\pdfoutput=1
\documentclass[twocolumn]{aastex61}

\usepackage{natbib,aas_macros} % Includes various useful things
\usepackage{hyperref}
\usepackage{graphicx}	% Including figure files
\usepackage{amsmath,amssymb} % maths part of aastex6
\usepackage{xspace}
\usepackage{enumerate}
\usepackage{fp} % variable calculations
\usepackage{lineno}
\usepackage{subfigure}
\shorttitle{Absolute Calibration of LMC NIR-TRGB}
\shortauthors{Hoyt, Freedman, \& Madore et al.}

\begin{document}


%% Old title
%Carnegie-Chicago Hubble Program: IV (??)  
%\\ An Absolute Calibration of the {JHK} and I-band 
% \\Tip of the Red Giant Branch Method \\.}

% New title
\title{The Near-Infrared Tip of the Red Giant Branch. II. \\
An Absolute Calibration in the Large Magellanic Cloud}


\correspondingauthor{Taylor J. Hoyt}\email{tjhoyt@uchicago.edu}

% principal
\author[0000-0001-9664-0560]{Taylor J. Hoyt}
\affiliation{Department of Astronomy \& Astrophysics, University of Chicago, 5640 South Ellis Avenue, Chicago, IL 60637}%\email{tjhoyt@uchicago.edu}

\author{Wendy L. Freedman} \affiliation{Department of Astronomy \& Astrophysics, University of Chicago, 5640 South Ellis Avenue, Chicago, IL 60637}
%\email{wfreedman@uchicago.edu}

\author[0000-0002-1576-1676]{Barry F. Madore}
\affil{Observatories of the Carnegie Institution for Science, 813 Santa Barbara St., Pasadena, CA~91101}
%\email{barry.f.madore@gmail.com}

\author{Mark Seibert}
\affil{Observatories of the Carnegie Institution for Science, 813 Santa Barbara St., Pasadena, CA~91101}
%\email{mseibert@obs.carnegiescience.edu}

% alphabetical 
\author[0000-0002-1691-8217]{Rachael L. Beaton}
\altaffiliation{Hubble Fellow}
\altaffiliation{Carnegie-Princeton Fellow}
\affiliation{Department of Astrophysical Sciences, Princeton University, 4 Ivy Lane, Princeton, NJ~08544}
%\affil{The Observatories of the Carnegie Institution for Science, 813 Santa Barbara St., Pasadena, CA ~~91101}

\author[0000-0003-2767-2379]{Dylan Hatt}
\affiliation{Department of Astronomy \& Astrophysics, University of Chicago, 5640 South Ellis Avenue, Chicago, IL 60637}
%\email{hatt@uchicago.edu}

\author[0000-0002-2502-0070]{In Sung Jang}
\affil{Leibniz-Institut fur Astrophysik Potsdam (AIP), An der Sternwarte 16, D-14482, Potsdam, Germany}
%\email{hanlbomi@gmail.com}

\author[0000-0003-2713-6744]{Myung Gyoon Lee}
\affil{Department of Physics and Astronomy, Seoul National University, Korea}
%\email{mglee@astro.snu.ac.kr}

\author{Andrew~J.~Monson}\affil{Department of Astronomy \& Astrophysics, Pennsylvania State University, 525 Davey Lab, University Park, PA 16802}
%
%\email{monson.andy@gmail.com}

\author{Jeffrey~A.~Rich}\affil{Observatories of the Carnegie Institution for Science, 813 Santa Barbara St., Pasadena, CA~91101}
%\email{jrich@obs.carnegiescience.edu}

%% are identified by the \altaffilmark after each name.  Specify  alternate

\begin{abstract} 

We present a new empirical \(JHK\) absolute calibration of the tip of the red giant branch (TRGB) in the Large Magellanic Cloud (LMC).  We use published data from the extensive \emph{Near-Infrared Synoptic Survey} containing 3.5 million stars, of which 65,000 are red giants that fall within one magnitude of the TRGB. Adopting the TRGB slopes from a companion study of the isolated dwarf galaxy IC\,1613 as well as an LMC distance modulus of \(\mu_0 = \)~18.49~mag from (geometric) detached eclipsing binaries, we derive absolute \(JHK\) zero-points for the near-infrared TRGB.  For comparison with measurements in the bar alone, we apply the calibrated \(JHK\) TRGB to a 500 deg\textsuperscript{2} area of the 2MASS survey. The TRGB reveals the 3-dimensional structure of the LMC with a tilt in the direction perpendicular to the major axis of the bar, in agreement with previous studies.

\end{abstract}

\keywords{galaxies: distances and redshifts; galaxies: individual: Large Magellanic Cloud; galaxies: stellar content; infrared: stars}
%\keywords{distances}

\section{Introduction}\label{sect:intro}

The Tip of the red giant branch (TRGB) method to determine extragalactic distances has been widely applied over the past two decades, primarily in the I-band.\footnote{ e.g., \citet[][]{lee93} \citet{riz07}; very recent applications of the I-band TRGB can be found in \citet{hatt17}, \citet{jang17}, and \citet{jang18cchp}.} This is, in part, due to the fortuitous convergence of peak $I$-band luminosities of first-ascent RGB stars as a function of color (or metallicity and age). Producing a luminosity function by marginalizing the CMD over color results in the sharpest discontinuity at the tip when viewed in the $I$-band, as opposed to bluer (or redder) bandpasses. In addition, TRGB stars are found in the halos of all types of galaxies (including spiral, irregular, face-on, edge-on, elliptical, and lenticular galaxies), as they all have a first generation of old (Population II) red giant branch stars. The ease with which the TRGB can be measured has translated into the publication of over 900 TRGB distances to some 302 individual galaxies (NED-D Release July 20, 2017).

Extending a calibration of the TRGB to the near-infrared (NIR) is of interest for many reasons. First and foremost, TRGB stars in the NIR are about 2 magnitudes brighter than in the I-band. At the same time the effects of line-of-sight extinction decrease at redder wavelengths. Furthermore, with the advent of larger, more sensitive telescopes in space, and especially given their performance in the near-infrared,  the NIR-TRGB method can be applied to more distant galaxies, thereby probing a larger cosmological volume. With a larger volume comes the immediate ability to calibrate additional Type Ia supernovae (SNe Ia) by obtaining TRGB distances to their host galaxies.

To date, there have been few empirical studies of the NIR TRGB. Using the Wide Field Camera 3 (WFC3) on the  Hubble Space Telescope (HST), \citet{dal12} obtained  NIR CMDs of 23 nearby galaxies, determining a single magnitude and representative color for each galaxy's TRGB. This study makes clear (in particular, see their Figure 21) the potential for empirically calibrating the slope of the NIR-TRGB. \cite{wu14} continue in a similar fashion, fitting the run of single TRGB magnitudes with representative colors for a large sample of galaxies. However, the NIR-TRGB slope is steep enough that a robust calibration requires a clear detection of, and fit to, the TRGB slope within a single galaxy.

In Paper I of this series on calibrating the NIR-TRGB, Madore et al. (2018) introduced a metallicity-independent method for accurately determining TRGB distances by rectifying, or flattening, the increase in NIR-TRGB magnitude with increasing NIR color. Slopes for the NIR-TRGB were determined as a function of \(JHK\) colors in the nearby, Local Group galaxy IC\,1613, where the TRGB loci are well-defined, the photometric precision is high, and potential systematic effects of reddening and photometric crowding/blending in images are minimized.

In this second paper, we undertake the next step and provide an absolute calibration for the NIR-TRGB using the Large Magellanic Cloud. Additionally we have used the precision of the NIR-TRGB method to probe the 3-D structure of the LMC. We describe the datasets used for this analysis in \autoref{sect:data}, provide the absolute calibration of the NIR-TRGB in \autoref{sect:calib}, and check the results of our calibration in \autoref{sect:geometry}.

\section{Multi-wavelength Observations \\of the Large Magellanic Cloud}\label{sect:data}

\begin{figure*} 
\centering 
\includegraphics[width=15.0cm]{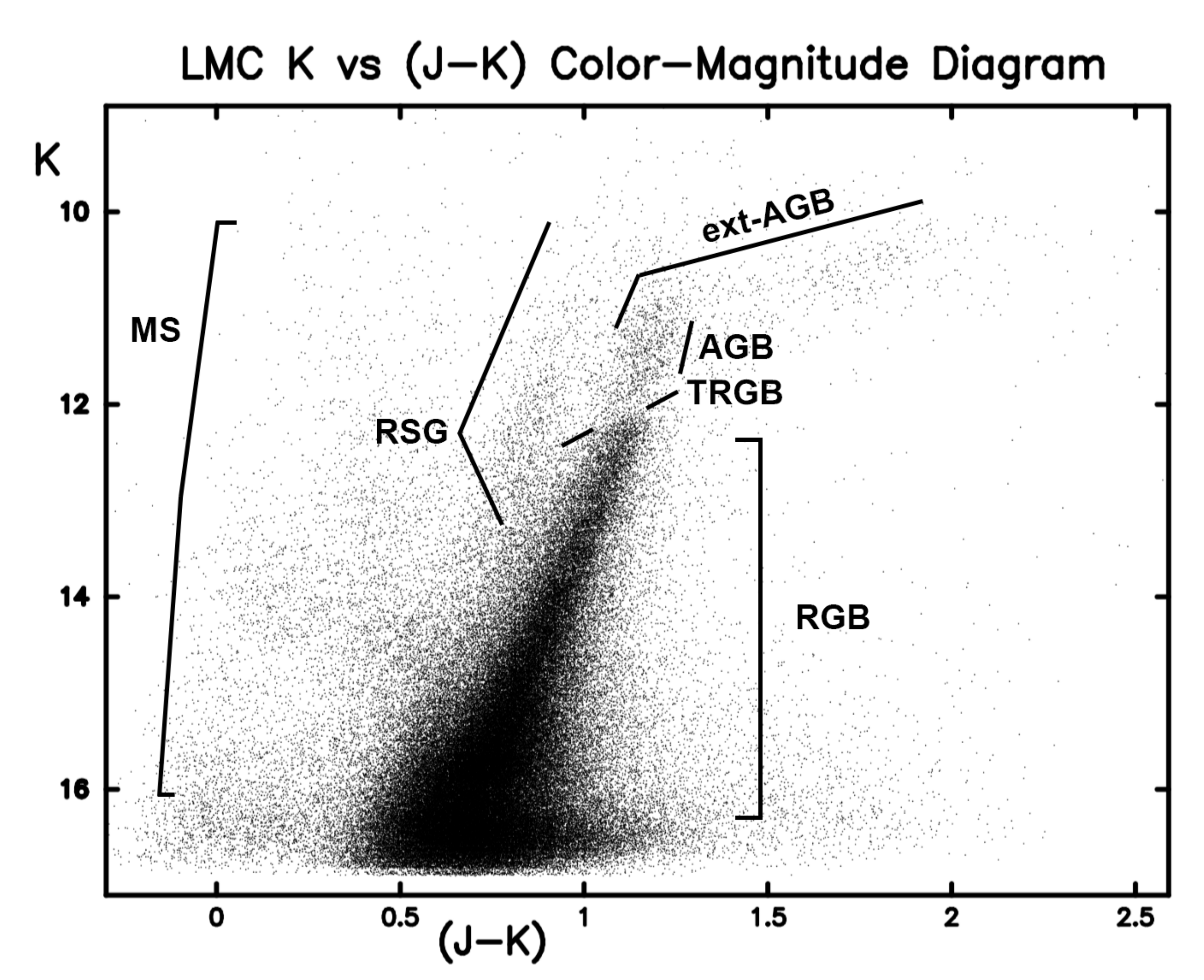} 
\caption{The $K$ vs $(J-K)$ color magnitude diagram for a subset of the 3.5 million LMC stars observed by \citet{mac15}. Only one star in ten is plotted so as to allow the slanting TRGB to be seen. Notable stellar components have been labeled.\label{fig:macri-CMD-JHK}
}
\end{figure*}

\subsection{The Near-Infrared Synoptic Survey}

We calibrate the zero points of the NIR-TRGB using publicly available $JHK$ photometry of stars in the central region of the LMC \citep{mac15}. The observations were taken as part of the Near Infrared Synoptic Survey (NISS) using the CPAPIR camera on the CTIO 1.5m telescope. The dataset includes \(JHK\) photometry of 3.5 million sources over $18$ deg$^2$ along the bar of the LMC to a limiting $K_s$ magnitude of 16.5 (see their Figure 1). 

\autoref{fig:macri-CMD-JHK} shows a representative sampling of the color-magnitude diagram (CMD) for the NISS data. Only $\sim$300,000 of the more than 3.5 million stars are shown, allowing the TRGB to be seen in contrast to its surrounding CMD components \citep[the entire sample can be seen in][their Figure 6]{mac15}. As can be readily appreciated, the NIR CMD is dominated by the red giant branch, rising up and tilting slightly to the red at a $(J-K)$ color between 0.8 and 1.2~mag. To give an indication of how populated the RGB luminosity function is near the TRGB, in the $K$ band we count 64,698 RGB stars in the first magnitude interval below the TRGB. The typical photometric error in all three bands at the expected TRGB magnitude is 0.005 mag.

The NISS dataset has a particular advantage for our purposes. Despite the observations being centered on the highest density region of the galaxy, stars in the main body of the LMC are likely less affected by the known tilt-induced, back-to-front spreading of the stars due to the LMC geometry and our particular viewing angle (\citealt{cal86, wel87}).

Accordingly, a determination of the distance to this centrally-located portion of the LMC is likely to be representative of the mean distance to the galaxy as a whole. Moreover, one of the highest accuracy determinations of the distance to the LMC \citep{pie13} is based on detached eclipsing binary (DEB) stars, most of which are in this same central region of the LMC. Furthermore, the kinematic center of the LMC and line of nodes fall along the elongated distribution of both the DEBs and the TRGB stars discussed in the first part of this paper. Thus, by performing our calibration in the LMC bar, we minimize geometric projection effects, which could otherwise blur or bias the TRGB distance determinations.

\begin{figure*} 
\centering
\includegraphics[width = \textwidth]{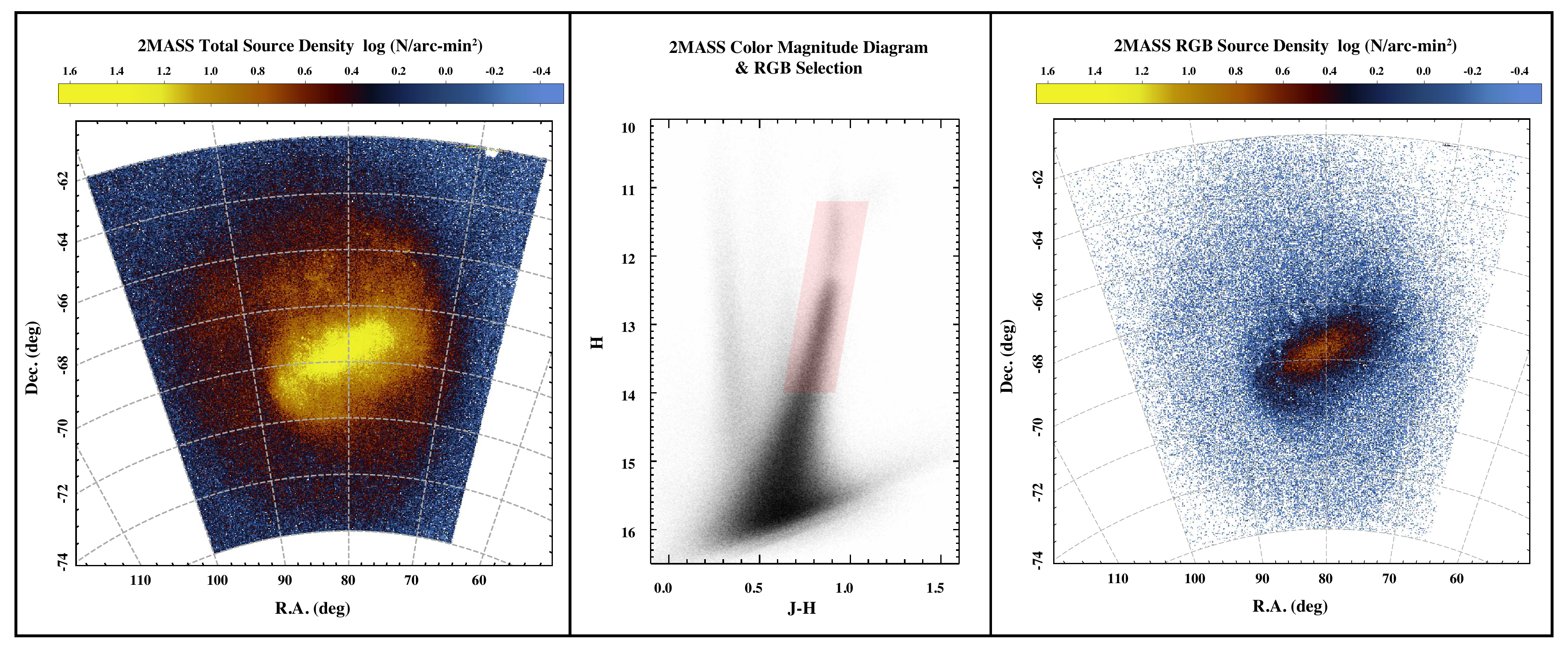}

\caption{(\emph{left}) Spatial distribution of stellar sources in the full catalog. The bar, disk, and spiral arms are apparent. (\emph{middle})  Color magnitude diagram of the 2 million stellar sources in the 2MASS LMC cutout that is discussed in the text. The red box encloses stellar sources near the visually apparent TRGB. (\emph{right}) Spatial distribution of stellar sources that lie within the red region of the CMD in the middle panel.}
\label{fig:2massall}
\end{figure*}

\subsection{The Two Micron All-Sky Survey}

To confirm that these geometric effects are indeed minimized, we query from the 2MASS All-Sky survey \citep{skr06} a \(35 \times 14 \) deg map of the LMC, for RA: 65-100 deg, DEC: -76 to -62 deg. This 2 million point source cutout is shown in panel (b) of \autoref{fig:2massall} and reveals a large density of sources outside the central region (bar) of the galaxy. Of these 2 million sources, 200,000 are stars within 1.5 mag of the TRGB, marked by a red region in panel (a) of \autoref{fig:2massall}. In panel (c) of the same figure is the spatial distribution of these stars located near the TRGB. \footnote{\citet{wei01} use an identical 2MASS cutout for their extensive study on various stellar populations found in the LMC. We refer the reader there for additional details on the 2MASS dataset used here.} We use this extended stellar distribution to determine if reddening and 3D projection effects within the bar are smaller than the errors of the presented calibration. 

Near the expected \(J\), \(H\), and, \(K\) TRGB magnitudes of 13.5, 12.5, and 12.5 mag, the median photometric errors are respectively, 0.025, 0.025, and 0.030 mag. Additionally, the \citet{mac15} data shares its photometric zero-point with 2MASS. As a result we avoid additional errors incurred by using different photometric systems.

\section{Calibrating The $JHK$ Tip of the Red Giant Branch}\label{sect:calib}

\subsection{Review of $JHK$ Slope Calibration in IC\,1613}\label{subsect:slopes}

Here we briefly summarize the slope calibration for the \(JHK\) magnitude-color relations as given in Paper I, which are based on observations of the Local Group dwarf galaxy IC\,1613 using \emph{FourStar} imaging at Las Campanas. The combination of large number statistics, high signal to noise photometry, and minimal contamination by younger stars makes our \emph{FourStar} IC 1613 dataset ideal for fitting the slope of the NIR-TRGB. Any dependence of NIR-TRGB magnitude on metallicity is being directly, and empirically, calibrated via the color of the TRGB stars.

First, the $JHK$ slopes in terms of $(J-K)$ colors: -0.85 $\pm$ 0.12, -1.62 $\pm$ 0.22; -1.85 $\pm$ 0.27.  Secondly, the $JHK$ slopes in terms of $(J-H)$ colors: -1.11 $\pm$ 0.15, -2.11 $\pm$ 0.26, -2.41 $\pm$ 0.36.

\subsection{LMC NIR-TRGB Zero-Points}
\label{subsect:zeropoints}

\begin{figure*} 
\centering 
\includegraphics[width=15.0cm]{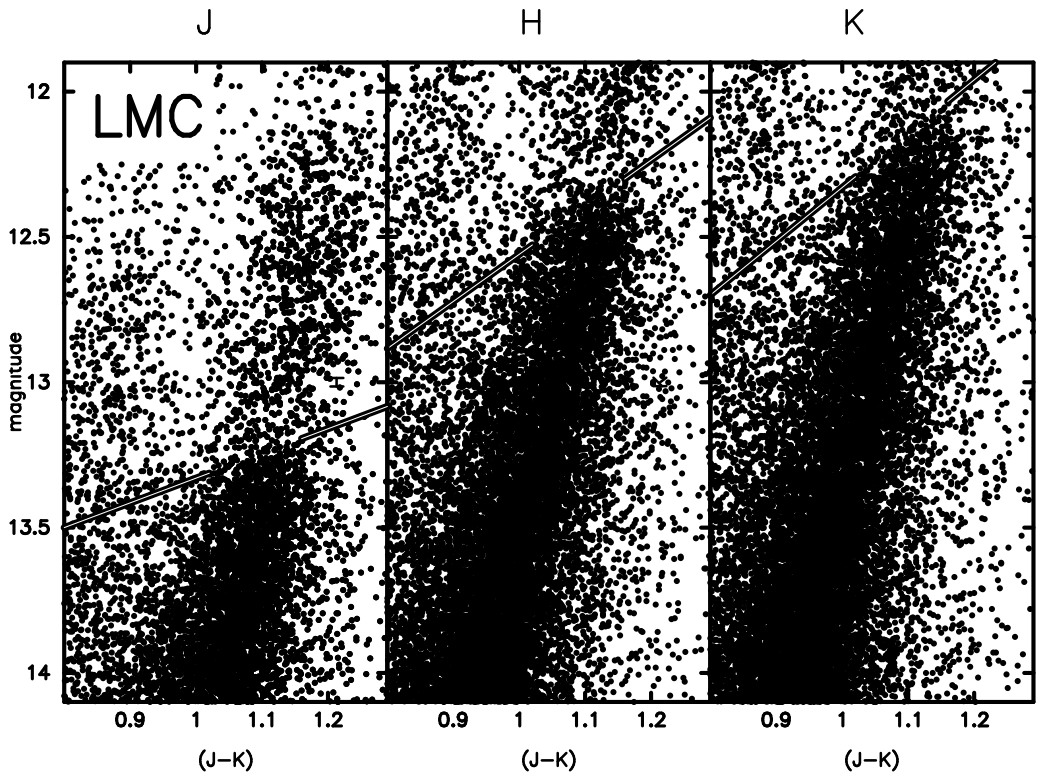} 
\caption{Zoomed-in versions of Fig. 1 focusing on the two magnitudes in $J$ (left panel) $H$ (middle panel) \& $K$ (right panel), each centered on the TRGB at $(J-K)$ = 0.8 to 1.3~mag. The slanting lines are fits to the slope of the TRGB discontinuity as determined in Paper I and briefly discussed in \autoref{sect:calib}.}
\label{fig:macri-JHK}
\end{figure*}

The method for determining the location of the TRGB using the slopes defined above is summarized in Paper I. In short, star magnitudes are rectified based on their colors (using the slopes defined in the previous section) so that the TRGB discontinuity is flat for a given band rather than slanted as would appear in the CMDs. The CMD is then marginalized over color to produce a luminosity function that is finely binned and smoothed in magnitude using Gaussian locally weighted regression (GLOESS). The magnitude at which the first derivative of the luminosity function peaks, as determined with a Sobel edge-detection kernel, is taken to be the magnitude of the TRGB.

\begin{figure*} 
\centering 
\includegraphics[width=8.5cm]{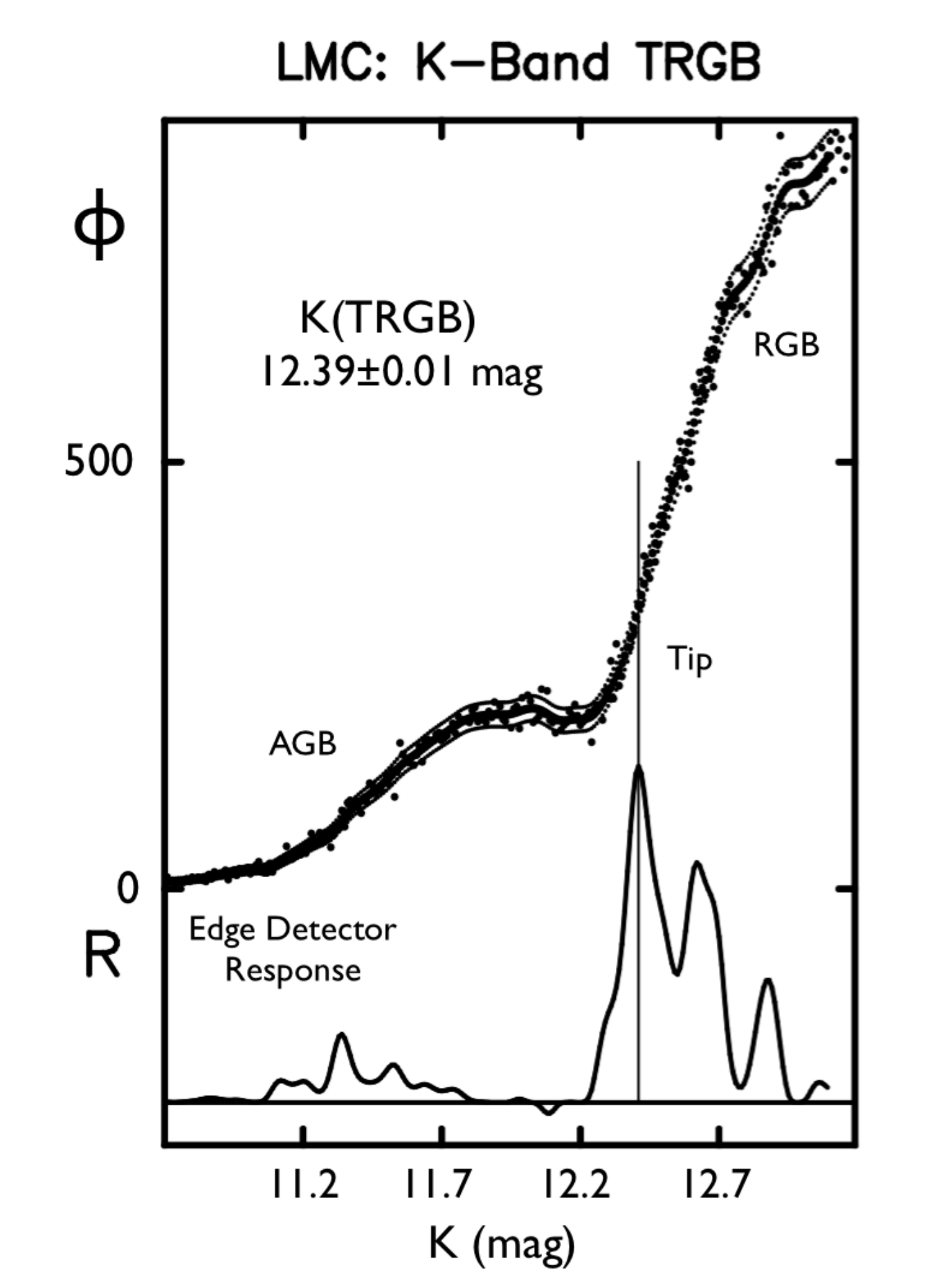}
\caption{GLOESS-smoothed K-band luminosity function for the stars in LMC drawn from the main (3.5 million star) sample of \citet{mac15}.  The luminosity function is produced by marginalizing over color in the \(K\) vs. \(J-K\) CMD. In the bottom portion of the panel, the Sobel filter edge-detector response peaks at $K$ = 12.39 $\pm$ 0.01~mag. Dots are the binned luminosity function, solid line is the smoothed LF, with 1-sigma Poisson boundaries marked by dashed lines. At the bottom of each, the TRGB is measured to be the maximum response in the Sobel filter output. The TRGB is marked by a vertical line extending from the response function up and into the luminosity function.
}
\label{fig:macridetect}
\end{figure*}

In \autoref{fig:macri-JHK} we show magnified portions of three color-magnitude diagrams of the NISS dataset in $JHK$ versus $(J-K)$, showing the two and a half magnitude range centered on the TRGB. The upward-slanting lines show the 2-dimensional fits to the NIR-TRGB discontinuities that were determined in Paper I. \autoref{fig:macridetect} shows the smoothed \(K\)-band luminosity function. The edge-detector response is shown in the lower portion of the figure. 

To assess and understand the errors in our TRGB detection, we have run a suite of simulations designed to cover a range of photometric errors, numbers of TRGB and AGB stars, and different widths of the Gaussian smoothing window. Given the large number of TRGB stars (65,000) one magnitude below the tip, the simulations proved insensitive to variations in these parameters and consistently yielded a statistical error no greater than $\pm$0.01 mag.

At this time, we adopt a distance modulus from DEBs to the LMC of 18.49 $\pm0.01$~mag (statistical) $\pm0.05$~mag (systematic) to provide our zero-point calibration. This value was determined by \citet{pie13}, based upon a weighted average of eight independent distances of DEBs in the bar of the LMC. We note also that this geometric distance modulus is consistent with the \emph{HST} parallax-based Cepheid distance to the LMC (18.48 $\pm$ 0.04~mag) determined jointly by \citet{mon12} and \citet{sco11}. 

To evaluate reddening effects on our calibration, we run our TRGB detection algorithm on a subsample of RGB stars selected to include only those which lie in regions of minimal reddening \( E(B-V) < 0.15\)~mag, as indicated by maps determined by \cite{nik04} using Cepheids and by \citet{pej09} using RR Lyrae variables. The result is a 0.02~mag brighter TRGB detection across all wavelengths. %Given the preliminary nature of this calibration, and pending a parallel effort which will be based on Gaia DR2 parallaxes of Milky Way TRGB stars
For this reason, we adopt an \(E(B-V) = 0.03 \pm 0.03\) ~mag. Given the results of our reddening test and that we do not expect RGB stars to be as reddened as Cepheids, this is a reasonable value and uncertainty. Adding the uncertainty on this adopted reddening in quadrature with the \citet{pie13} systematic error (for which the reddening uncertainties are estimated to be $\sim$0.4\%) brings our total systematic error to $\pm$0.06 mag.

First, the resulting calibrations in terms of $(J-K)$ colors:
\begin{align}
M_J = -5.14 - 0.85 \left[(J-K) - 1.00\right] \\ 
M_H = -5.94 - 1.62 \left[(J-K) - 1.00\right] \\ 
M_K = -6.14 - 1.85 \left[(J-K) - 1.00\right]  
\end{align}
\noindent
Errors on the slopes are \(\pm\) 0.12, 0.22 and 0.27 in J, H \& K, respectively. Secondly, the equivalent calibrations in terms of $(J-H)$ colors:
\begin{align}
M_J = -5.13 - 1.11 \left[(J-H) - 0.80\right] \\ 
M_H = -5.93 - 2.11 \left[(J-H) - 0.80\right] \\ 
M_K = -6.13 - 2.41 \left[(J-H) - 0.80\right]  
\end{align}
\noindent Errors on these slopes are \(\pm\) 0.15, 0.26 and 0.36 in J, H \& K, respectively.

We adopt for the systematic error on the zero points the \citet{pie13} detached eclipsing binary systematic error added in quadrature to the error on our adopted reddening value. The error on the NIR-TRGB zero point calibration is thus $\pm0.01$~mag (statistical) $\pm0.06$~mag (systematic).

\subsection{NIR-TRGB Zero-Point Comparisons} \label{subsect:compare}

We have presented above a calibration of the zero-point of the TRGB at three ($JHK$) near-infrared wavelengths, where the color coefficients come from a companion study of the dwarf galaxy IC\,1613 (Paper I). The zero-points, derived in this paper, come from the detached eclipsing variable star parallaxes \citep{pie13}, tied directly to TRGB stars in the same central bar region of the LMC. 

We note that our calibration is in good agreement with recent theoretical work. \citet{ser17} provide 2MASS \(J\) and \(K\) TRGB equations for two color regimes. Shifting their \( (J-K) > 0.76 \) equations to our fiducial \(J-K = 1.00\) pivot, we see that their K-band TRGB has a zero point and slope of -6.17 and -1.811, to be directly compared to the calibration in the previous section, and in \autoref{tab:calcomp}. The theoretical slopes and zero points are well within the errors of our empirical ones. This agreement with theory is certainly promising; nevertheless our distance-scale calibration remains empirical.

\begin{deluxetable}{clll}
\tablenum{1}

\tablehead{ \colhead{K-band Calibration} & \colhead{\(M_K\)} & \colhead{ Slope } & \colhead{Notes} }

\startdata
\citet{gor16}         & \( -6.32 \) & \( -2.15 \) & Globular Clusters \\
\citet{ser17}         & \( -6.17 \) & \(-1.811 \) & Theory \\
\hline
\textbf{This study} & \( -6.14 \) & \(-1.85 \) & IC 1613 and LMC\\
%MTRGB(F160W) =?2.576(F110W ?F160W)?3.496 (1)
\enddata
\caption{\small Comparison of recent K-band TRGB calibrations to that provided in this study, all in the 2MASS photometric system. In the Appendix, we describe the mapping of the \citet{gor16} metallicity calibration onto the empirical \(K\)-\((J-K)\) plane. }
\label{tab:calcomp}
\end{deluxetable}

A recent paper by \citet{gor16} draws very different conclusions about the NIR-TRGB zero-points, as shown in \autoref{tab:calcomp}. \citet{gor16} attempt to fit a metallicity dependence to the NIR-TRGB slope, in contrast with the purely empirical calibration described in Paper I. Our approach is to use directly observable colors, such as $(J-K)$, avoiding other transformations, bolometric corrections or any explicit dependencies on theory, and to use an extensive catalog of LMC stars to define the zero point of the calibration. As \cite{ser17} emphasize, there are $\sim$0.2 mag uncertainties in bolometric corrections for the theoretical TRGB; in addition, small-number statistics for the 24 calibrating globular clusters used by \citet{gor16} may preclude a robust and accurate detection of the TRGB.  Hence, our preference is to adopt an empirical TRGB calibration, and to limit the calibration to systems where small-number statistics are not a limiting issue. 

\section{3-dimensional LMC Geometry}\label{sect:geometry}

In this section we compare the NIR-TRGB in an extended spatial sample of the LMC to an inner disk/bar region. To this end we break up the TRGB-selected source catalog from \autoref{fig:2massall}, panel (c), into 39 bins of equal source count. This is intended to keep constant the signal to noise of each field's TRGB discontinuity. 

\begin{figure*}
\includegraphics[width=\textwidth]{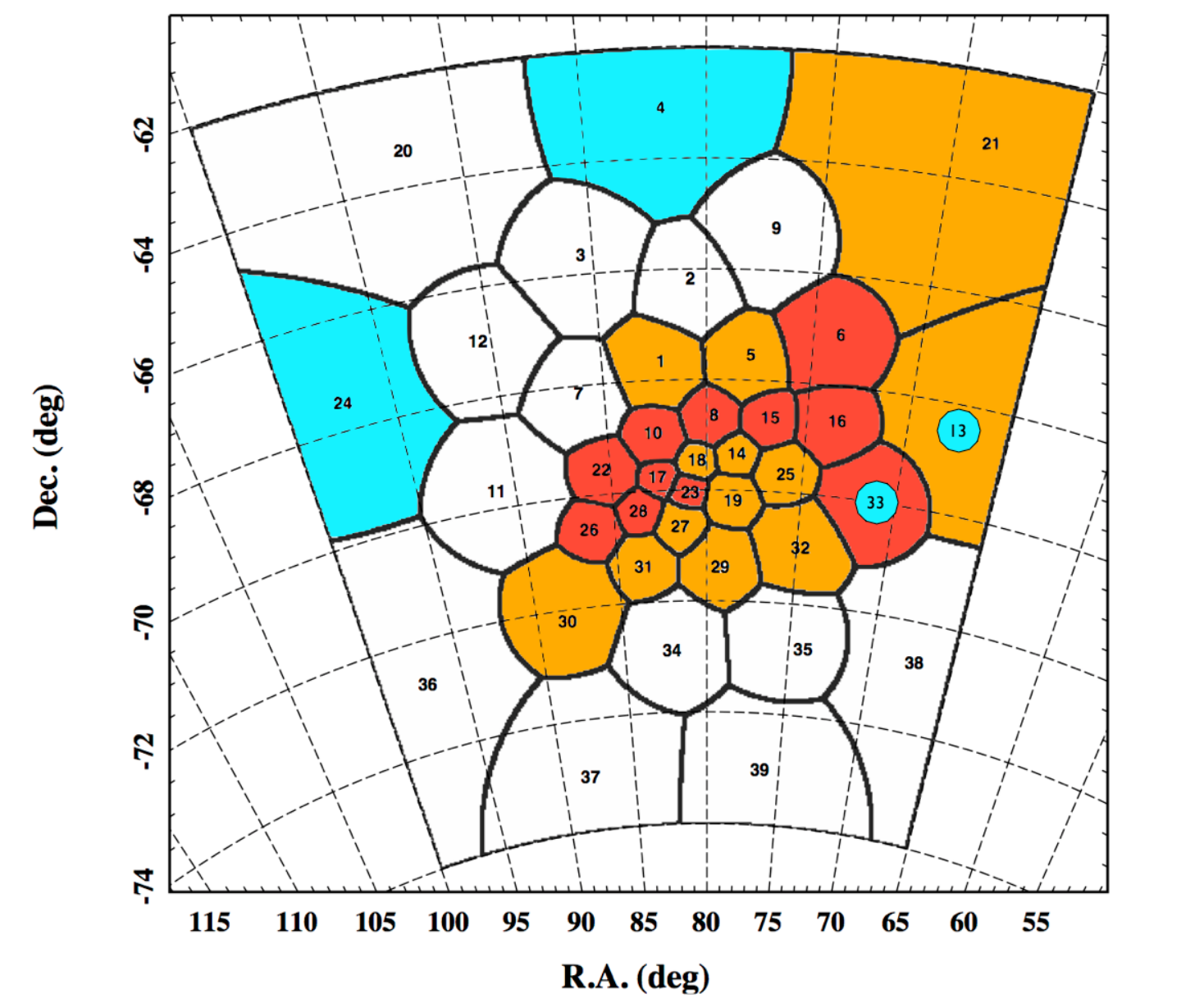}
\caption{Voronoi tesselation results. Each field is numbered corresponding to its entry number in \autoref{tab:2massdists}. The orange color marks regions selected to represent the LMC bar and line of nodes region in the 2MASS dataset. Red denotes candidate fields that were excluded from the bar region based on the crowding and reddening criteria discussed in the text. Lastly, blue denotes a field with its example CMD, luminosity function, and response function displayed in \autoref{fig:fieldcmds}.
}
\label{fig:vorobounds}
\end{figure*}

The bins are generated via an adaptive voronoi binning technique laid out by \citet{die06}, which is a generalization of the \citet{cap03} voronoi binning algorithm. The algorithm starts at one ``pixel,'' in this case a stellar source position, and continues to absorb the nearest pixel until the bin contains approximately 5700 stellar sources. The process continues with the nearest unbinned pixel under the constraints that i) there are no holes or overlapping bins, ii) the bins satisfy a pre-defined Roundness criteria, and iii) that the scatter in stellar counts in each bin is minimized. The results of the voronoi tesselation are shown in \autoref{fig:vorobounds}.

\begin{figure*}
\centering
\includegraphics[width = 0.475\textwidth, trim={0 0.3cm 0.3cm 0.2cm},clip ]{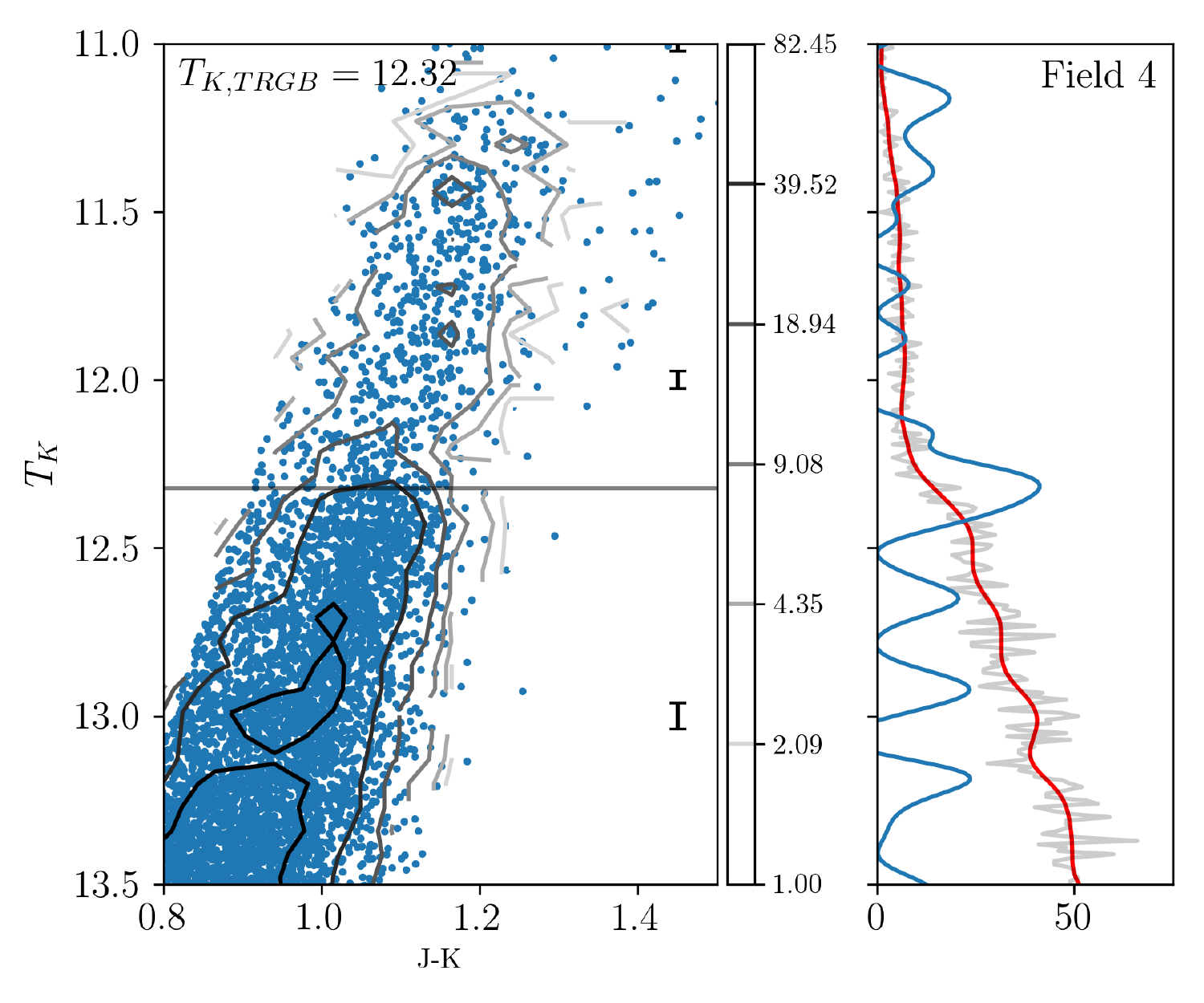}
\includegraphics[width = 0.475\textwidth, trim={0 0.3cm 0.3cm 0.2cm},clip ]{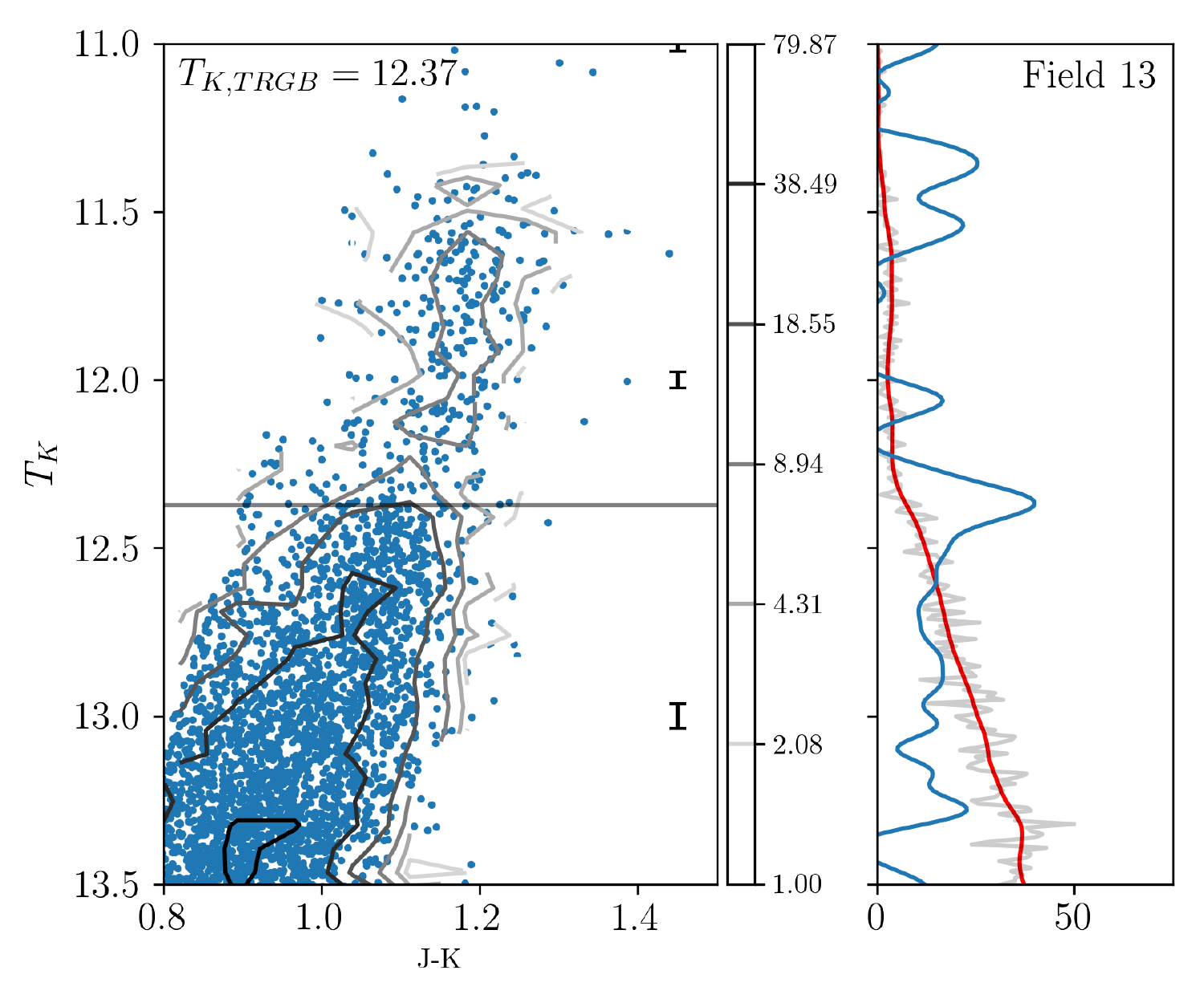}
\includegraphics[width = 0.475\textwidth, trim={0 0.3cm 0.3cm 0.2cm},clip ]{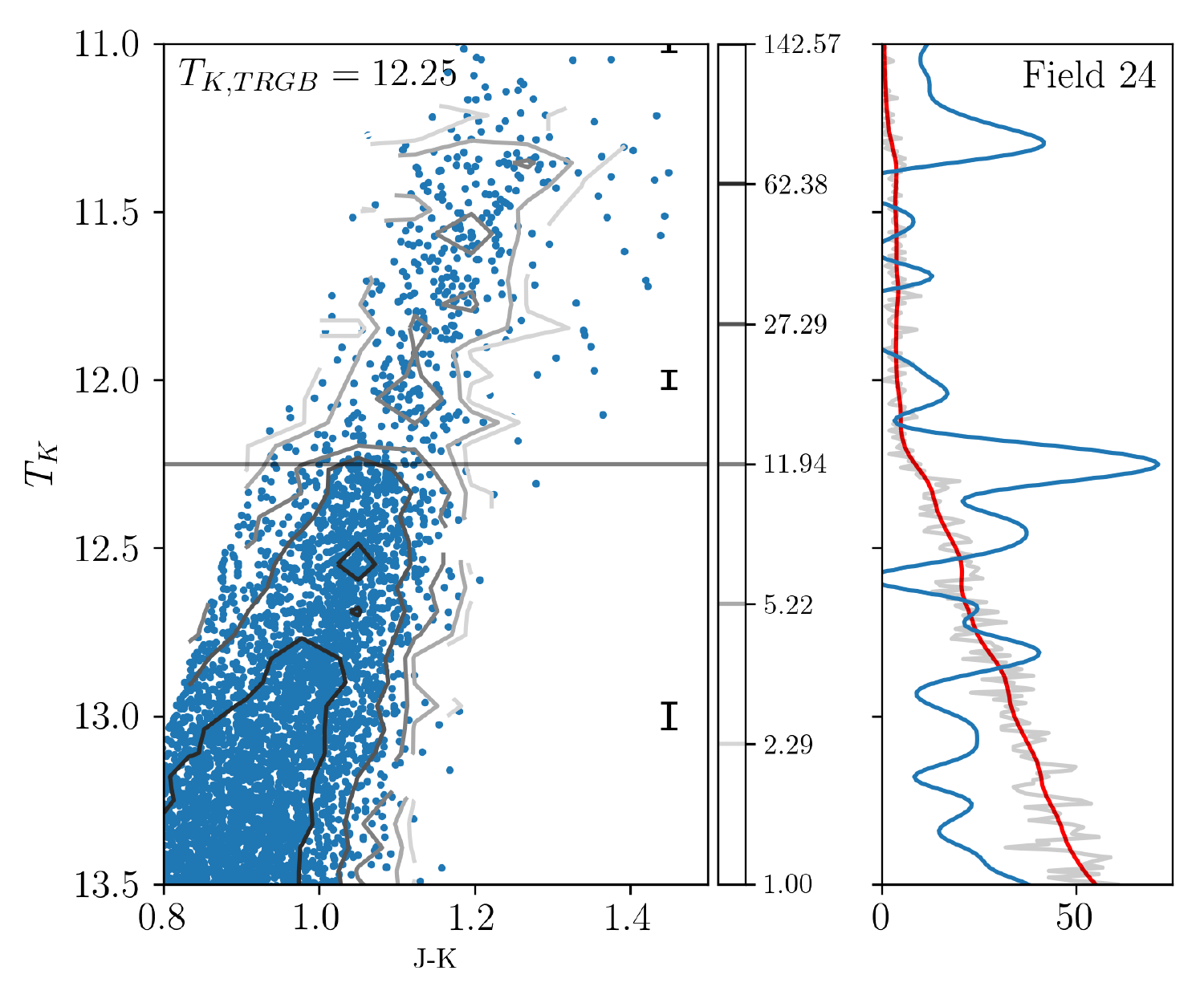}
\includegraphics[width = 0.475\textwidth, trim={0 0.3cm 0.3cm 0.2cm},clip ]{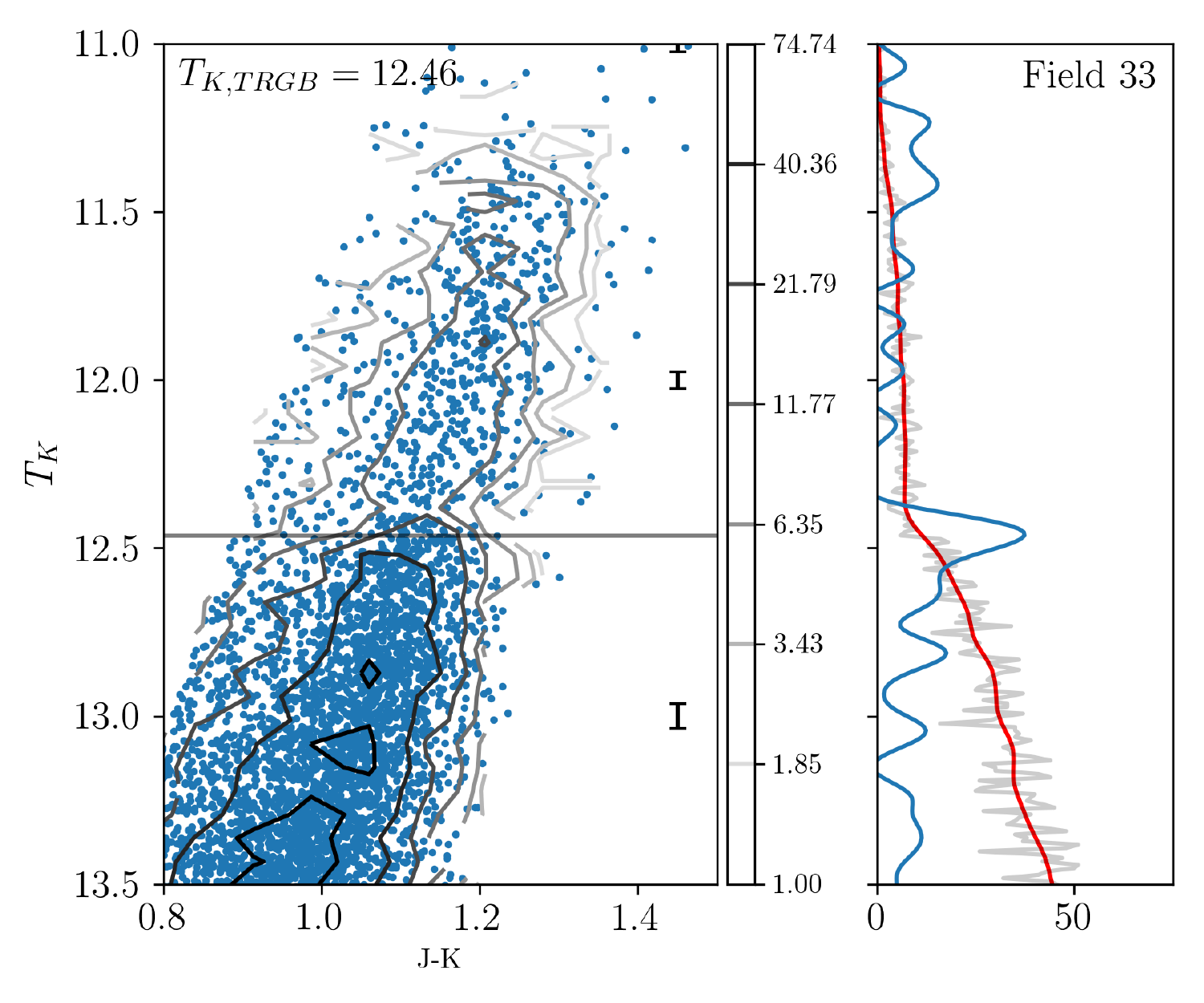}
\caption{Four representative detections of the rectified K-band \(T_K\) vs. \(J-K\) for 2MASS fields 4, 13, 24, and 33. The plotted contours are for bins of dimension \( \Delta (J-K) = 0.05\mathrm{~mag}, \Delta T_K = 0.08\mathrm{~mag} \). The contour color scaling is located to the right of each CMD. The CMD color cuts were determined using the region in the 2MASS data outside RA: [68, 95] deg and Dec: [-74, -64] deg. Error bars in the CMDs represent median magnitude errors. The right panel of each plot show a finely (0.005 mag) binned luminosity function in gray, the smoothed luminosity function in red, and the edge detection response function in blue. The point of maximum edge detection response is displayed in each CMD as a horizontal line.}
\label{fig:fieldcmds}
\end{figure*}

\begin{figure*}
\centering
\includegraphics[width = \textwidth,  trim={2cm 8cm 2cm 8cm},clip ]{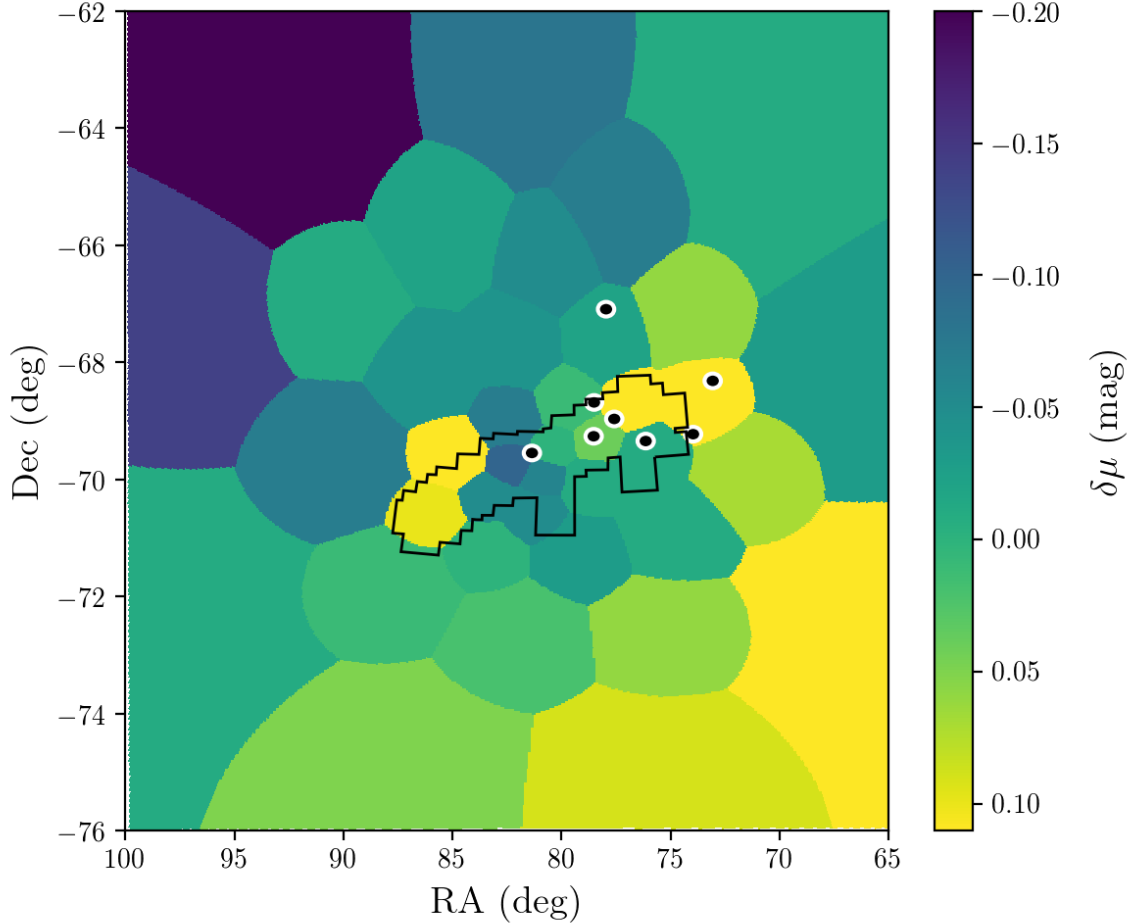}
\caption{Map of differential distance moduli to each field in the 2MASS cutout defined to be \( \delta \mu =  \mu_i - \mu_{cal} \) where \(\mu_i\) is the distance to any one 2MASS field and \(\mu_{cal} = 18.49 \) mag is the \citet{pie13} DEB distance adopted to perform the absolute calibration in the preceding section. To ensure that the TRGB is sufficiently populated the total number of stars near the TRGB in each field is held fixed. The color of each field corresponds to the weighted mean of the NIR-TRGB distances provided in \autoref{tab:2massdists}. The \citet{mac15} region is contained in the black rectangle. The eight eclipsing binaries of \citet{pie13} are plotted as white-circled black points. We see a remarkably clear tilt along the NE-SW direction and minimal variation along the direction parallel to the major axis of the bar, consistent with the bar being aligned with the line of nodes.}
\label{fig:2massmap}
\end{figure*}

\begin{figure*}
\centering
\includegraphics[]{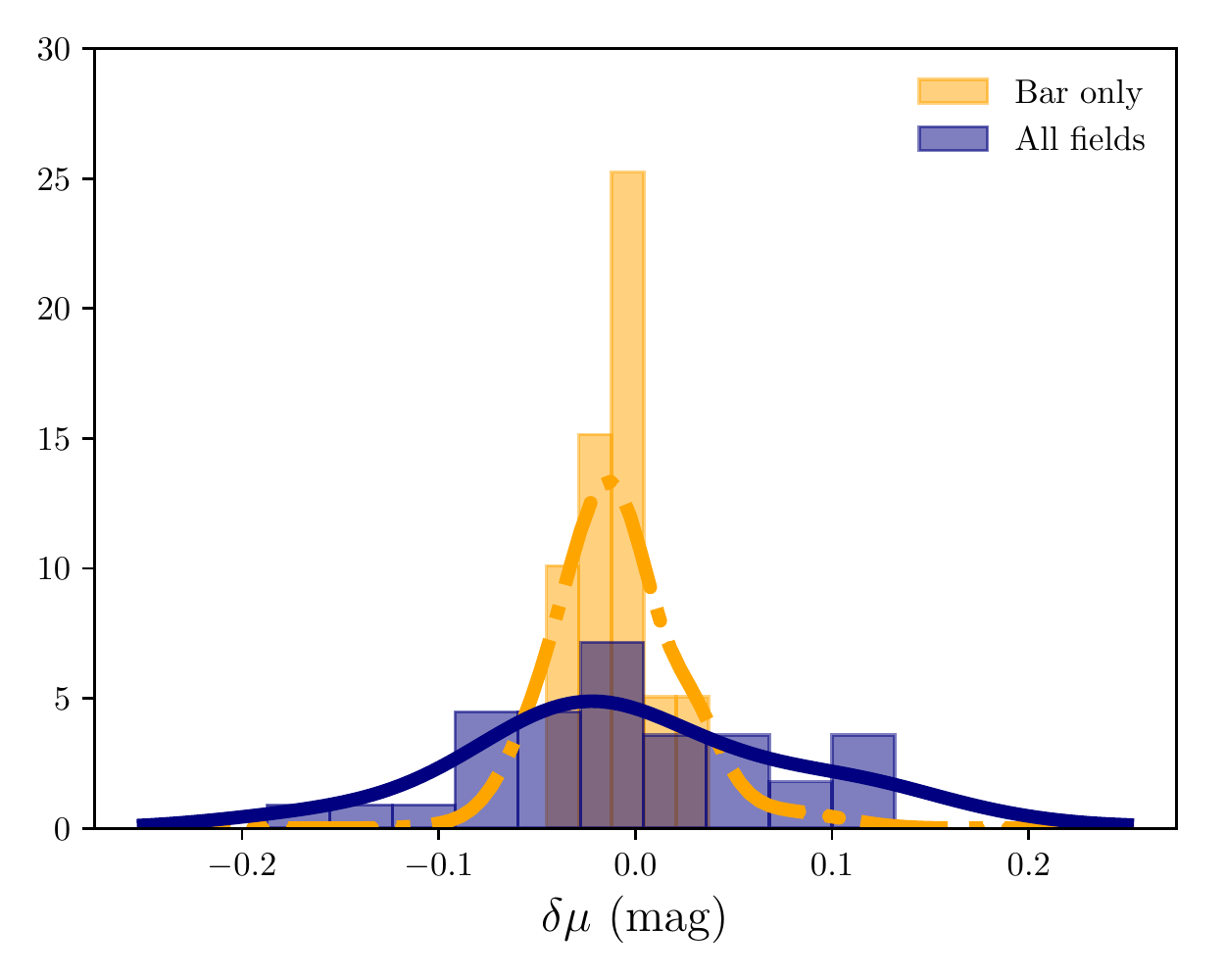}
\caption{\small Normalized histograms, both smoothed and unsmoothed, of differential distance moduli \(\delta \mu\) to each 2MASS LMC field where \( \delta \mu =  \mu_i - \mu_{cal} \), \(\mu_i\) is the distance to any one 2MASS field and \( \mu_{cal} = 18.49 \) mag is the \citet{pie13} DEB distance adopted to perform the absolute calibration in the preceding section. The distance modulus distribution for the 2MASS bar region, defined in the text and marked in \autoref{fig:vorobounds}, is plotted in dot-dashed orange. The mean of this bar distribution is -0.01 mag brighter than the adopted DEB distance modulus, with a dispersion of 0.02 mag. The distribution of distances to all 39 2MASS fields is plotted in solid navy. The mean of the full distribution is also -0.01 mag brighter than the adopted distance modulus with a dispersion of 0.07 mag. All distributions, smoothed and unsmoothed, are normalized to unity.}
\label{fig:2masshists}
\end{figure*}

To assess 3-D projection effects on our calibration, we apply the calibration from \autoref{sect:calib} to measure TRGB distances to each field. We measure the TRGB individually in \(H\) and \(K\) for both \(J-H\) and \(J-K\) colors.\footnote{We omit J because, by construction, the rectified \(T_J\) magnitude is exactly equal to \(T_H\) and \(T_K\) for colors \(J-H\) and \(J-K\), respectively.} \autoref{fig:fieldcmds} shows example detections of the NIR-TRGB in four of the 39 LMC fields (colored blue in \autoref{fig:vorobounds}). The distances to each of the 39 fields are provided in \autoref{tab:2massdists}, in map form in \autoref{fig:2massmap}, and as smoothed histograms in \autoref{fig:2masshists}.

In Figure 6 we see a clear tilt along the direction perpendicular to the major axis of the bar and the expected line of nodes, running closer (18.3 mag) in the Northeast to more distant (18.6 mag) in the Southwest. The minimal variation, \(\sigma<0.02\) mag along the NW-SE direction provides further evidence that the bar's major axis is aligned with the line of nodes. This is in agreement with previous studies, first by \cite{dev72}, then using Cepheids \citep[e.g.,][]{cal86,wel87,per04,inn16}, carbon stars \citep{wei01, mar01, ols11}, RR Lyrae \citep{deb14}, and the red clump \citep{sub13}. That we have backed out this known geometry is a testament to the potency of the present NIR-TRGB calibration.

Though an in depth analysis of the LMC geometry is beyond the scope of this calibration paper, the NIR-TRGB has allowed us to probe a greater, contiguous spatial extent of the LMC than previous studies. As such, we briefly compare qualitatively our results to those from studies with a similar spatial extent. \citet{inn16} determine optical distances to a handful of Cepheids in the SW corner of the LMC, contained within RA: [-78,-68] deg, Dec: [-75,-73] deg. They find that these Cepheids are all 0.1-0.15 mag fainter than the mean LMC distance, in agreement with our measurements. On the other hand, we see less agreement with the results of \citet{wei01} using Carbon Stars (their Fig. 11). We see a \(\sim\)0.1 mag brightening in the measured distance modulus running from the northern to the southern fields, in contrast with their null result. Inversely, we see less of a drift, approx. 0.15 mag, running from the west to the east as opposed to their measured 0.3 mag drift. This is likely down to the western carbon stars lying in exceptionally reddened regions, a scenario evidenced by our full differential distance map as well as later reddening maps \citep[e.g.][]{nik04,pej09,inn16}.

To represent the bar of the LMC we choose 13 of the 39 voronoi regions from the inner region of the 2MASS map. These fields are chosen to avoid contamination by red supergiants (RSGs) and early-AGB stars (marked red in \autoref{fig:vorobounds}).\footnote{\citet{wei01} used 2MASS data of the LMC to trace this population of stars and we use their maps to make the selections. We exclude from our bar analysis regions in which the number density of these young, red stars exceeds \(120 \mathrm{~deg}^{-2}\).} Those fields selected to be part of this uncontaminated 2MASS bar region are marked with asterisks in \autoref{tab:2massdists} and are colored gold in \autoref{fig:vorobounds}. This 2MASS bar region is used to confirm that the results of each of our \(JHK\) calibrations agree with one another, as well as with the adopted DEB distance.

As seen in \autoref{fig:2masshists}, there is a small 0.01 mag offset in the mean of the 2MASS distance distribution from the adopted distance of 18.49 mag, suggesting that 3D projection effects and, in the NIR, extinction effects are small compared to the precision of the metallicity-independent calibration provided here. %Finally, it can be confirmed in \autoref{tab:2massdists} that all four independent filter-color combinations agree to within 0.01 mag in mean distance to the LMC.

We briefly comment on reddening here. \citet{nik04} combine theoretical pulsation models of Cepheids, star formation history reddenings with \emph{HST}, and low-reddening Cepheids to compute a reddening map of the LMC inner disk (their Figure 9). In their map is the notable reddening spike at RA \( \simeq 86^\circ \), Dec \(\simeq 69^\circ \), which is consistent with reddening maps constructed by and \citet{inn16} also with Cepheids, and \citet{pej09} using RR Lyrae.  This region exactly coincides with Fields 22 and 26 in our 2MASS map. As can be seen in \autoref{fig:2massmap} these two fields are \(\sim\) 0.1 mag fainter than the surrounding neighbors, consistent with the \citet{nik04} \(E(B-V)\) values of 0.4-0.5~mag in that region.

In \autoref{fig:2masshists} we present histograms of distance to each field in both the 2MASS bar region and the full 2MASS dataset. For the bar region, the dispersion in distance modulus is 0.02 mag and the mean is 18.48 mag, only 0.01 mag fainter than the adopted distance of 18.49 mag -- well within the uncertainties of the calibration. Panel (b) includes the remaining 26 fields and the scatter increases by roughly a factor of three. Importantly the mean does not shift, indicating that the source of this increased scatter is not a one-way effect like reddening, but rather an effect symmetric in apparent distance modulus e.g. an observable tilt in the LMC.\footnote{A reddening gradient is not likely to produce the distribution of distances observed. The two most distant regions at the NE and SW corners are located at the same radii from the LMC center, but their distances, as determined in the NIR, differ by more than 0.3 mag.} Note the small asymmetric bump in the full histogram around \(\mu \sim 18.6\) mag. This bump is sourced by highly reddened fields, in particular the fields 22 and 26 discussed earlier. The strong presence of dust is consistent with these reddened fields satisfying our RSG density cutoff of \(>\)120 RSGs/deg\(^2\) for the 2MASS bar sample, being regions of high star formation.

\begin{deluxetable*}{ccccccccccccc}

%% Over-ride the default font size
%% Use 8pt font
\tabletypesize{\small}

%% Use \tablewidth{?pt} to over-ride the default table width.
%% If you are unhappy with the default look at the end of the
%% *.log file to see what the default was set at before adjusting
%% this value.

%% This is the title of the table.
\tablecaption{}

\tablenum{2}

%% The \tablehead gives provides the column headers.  It
%% is currently set up so that the column labels are on the
%% top line and the units surrounded by ()s are in the 
%% bottom line.  You may add more header information by writing
%% another line between these lines. For each column that requries
%% extra information be sure to include a \colhead{text} command
%% and remember to end any extra lines with \\ and include the 
%% correct number of &s.
\tablehead{\colhead{Field no.} & \colhead{\(\mu_{H, J-H}\)} & \colhead{\(\delta \mu_{H, J-H}\)} & \colhead{\(\mu_{H, J-H}\)} & \colhead{\(\delta \mu_{H, J-K}\)} & \colhead{\(\mu_{H, J-H}\)} & \colhead{\(\delta \mu_{K, J-H}\)} & \colhead{\(\mu_{H, J-H}\)} & \colhead{\(\delta \mu_{K, J-K}\)} & \colhead{\(\mu_{avg}\)} & \colhead{\(\delta \mu_{avg}\)} & \colhead{\(\Delta\)RA} & \colhead{\(\Delta\)Dec} \\ 
\colhead{} & \colhead{(mag)} & \colhead{(mag)} & \colhead{(mag)} & \colhead{(mag)} & \colhead{(mag)} & \colhead{(mag)} & \colhead{(mag)} & \colhead{(mag)} & \colhead{(mag)} & \colhead{(mag)} & \colhead{(deg)} & \colhead{(deg)}
} 

%% All data must appear between the \startdata and \enddata commands
\startdata
1* & 18.44 & 0.02 & 18.48 & 0.06 & 18.47 & 0.10 & 18.46 & 0.03 & 18.45 & 0.02 & 1.52292 & 2.05056 \\
2 & 18.46 & 0.04 & 18.39 & 0.08 & 18.44 & 0.04 & 18.43 & 0.08 & 18.44 & 0.03 & -0.01875 & 3.62278 \\
3 & 18.48 & 0.10 & 18.40 & 0.10 & 18.49 & 0.10 & 18.48 & 0.05 & 18.47 & 0.04 & 4.79375 & 3.99778 \\
4 & 18.41 & 0.10 & 18.40 & 0.05 & 18.44 & 0.10 & 18.42 & 0.05 & 18.41 & 0.03 & 0.73125 & 6.43944 \\
5* & 18.46 & 0.10 & 18.48 & 0.10 & 18.47 & 0.10 & 18.47 & 0.07 & 18.47 & 0.04 & -2.81042 & 2.16861\\
6 & 18.56 & 0.10 & 18.66 & 0.15 & 18.48 & 0.10 & 18.55 & 0.04 & 18.55 & 0.03 & -7.01875 & 2.64778 \\
7 & 18.45 & 0.06 & 18.47 & 0.10 & 18.47 & 0.10 & 18.42 & 0.10 & 18.45 & 0.04 & 5.23125 & 1.61167 \\
8 & 18.61 & 0.20 & 18.43 & 0.15 & 18.56 & 0.30 & 18.46 & 0.30 & 18.50 & 0.10 & -1.08125 & 1.10611\\
9 & 18.42 & 0.05 & 18.43 & 0.08 & 18.36 & 0.15 & 18.44 & 0.10 & 18.42 & 0.04 & -3.89375 & 4.50194\\
10 & 18.45 & 0.10 & 18.43 & 0.10 & 18.41 & 0.10 & 18.39 & 0.10 & 18.42 & 0.05 & 1.71042 & 0.76444\\
11 & 18.43 & 0.10 & 18.42 & 0.10 & 18.42 & 0.10 & 18.42 & 0.10 & 18.42 & 0.05 & 9.98125 & 0.08944 \\
12 & 18.63 & 0.20 & 18.41 & 0.10 & 18.68 & 0.20 & 18.44 & 0.20 & 18.48 & 0.08 & 9.48125 & 2.75611 \\
13* & 18.47 & 0.05 & 18.44 & 0.06 & 18.48 & 0.08 & 18.46 & 0.04 & 18.46 & 0.03 & -13.01880 & 1.75611 \\
14* & 18.51 & 0.10 & 18.53 & 0.10 & 18.56 & 0.15 & 18.52 & 0.10 & 18.53 & 0.05 & -2.51875 & 0.40333\\
15 & 18.67 & 0.20 & 18.77 & 0.20 & 18.63 & 0.20 & 18.50 & 0.20 & 18.64 & 0.10 & -4.14375 & 1.08528 \\
16 & 18.61 & 0.15 & 18.60 & 0.10 & 18.61 & 0.08 & 18.60 & 0.10 & 18.61 & 0.05 & -7.39375 & 1.08944\\
17 & 18.38 & 0.05 & 18.42 & 0.10 & 18.33 & 0.07 & 18.41 & 0.05 & 18.39 & 0.03 & 1.66875 & -0.00083\\
18* & \ldots & \ldots & 18.47 & 0.15 & 18.51 & 0.10 & 18.48 & 0.15 & 18.49 & 0.07 & -0.45625 & 0.29778 \\
19* & 18.51 & 0.08 & 18.45 & 0.08 & 18.50 & 0.07 & 18.48 & 0.05 & 18.48 & 0.03 & -2.45625 & -0.43556 \\
20 & 18.37 & 0.15 & 18.33 & 0.15 & 18.28 & 0.04 & 18.30 & 0.04 & 18.29 & 0.03 & 12.23130 & 5.88944\\
21* & 18.47 & 0.05 & 18.46 & 0.06 & 18.52 & 0.10 & 18.49 & 0.05 & 18.48 & 0.03 & -10.89380 & 5.75611\\
22 & 18.61 & 0.20 & 18.65 & 0.20 & 18.59 & 0.20 & 18.62 & 0.20 & 18.62 & 0.10 & 4.48125 & 0.17000 \\
23 & 18.45 & 0.06 & 18.44 & 0.06 & 18.39 & 0.08 & 18.43 & 0.05 & 18.43 & 0.03 & -0.01875 & -0.26056\\
24 & 18.36 & 0.04 & 18.30 & 0.10 & 18.33 & 0.05 & 18.35 & 0.04 & 18.35 & 0.02 & 15.60620 & 2.24778\\
25* & 18.48 & 0.06 & 18.48 & 0.08 & 18.49 & 0.10 & 18.47 & 0.07 & 18.48 & 0.04 & 5.10208 & 0.09778\\
26 & 18.60 & 0.10 & 18.56 & 0.15 & 18.59 & 0.15 & 18.60 & 0.10 & 18.59 & 0.06 & 5.43958 & -0.87722 \\
27* & 18.44 & 0.05 & 18.46 & 0.10 & 18.42 & 0.08 & 18.47 & 0.10 & 18.44 & 0.04 & 0.52292 & -0.85500\\
28 & \ldots & \ldots & 18.34 & 0.10 & \ldots & \ldots & 18.54 & 0.10 & 18.44 & 0.07 & 2.66875 & -0.59389\\
29* & 18.51 & 0.08 & 18.49 & 0.08 & 18.47 & 0.04 & 18.45 & 0.03 & 18.46 & 0.02 & -1.76875 & -1.64667\\
30* & 18.50 & 0.05 & 18.48 & 0.10 & 18.51 & 0.05 & 18.49 & 0.10 & 18.50 & 0.03 & 7.35625 & -2.26056\\
31* & 18.44 & 0.10 & 18.43 & 0.10 & 18.58 & 0.10 & 18.52 & 0.15 & 18.49 & 0.05 & 2.58542 & -1.59528\\
32* & 18.48 & 0.05 & 18.48 & 0.04 & 18.49 & 0.06 & 18.49 & 0.06 & 18.48 & 0.03 & -6.14375 & -1.27306\\
33 & 18.54 & 0.04 & 18.57 & 0.03 & 18.63 & 0.15 & 18.56 & 0.03 & 18.56 & 0.02 & -10.14380 & -0.24389\\
34 & 18.52 & 0.07 & 18.52 & 0.08 & 18.50 & 0.08 & 18.51 & 0.08 & 18.51 & 0.04 & 0.85625 & -3.16889\\
35 & 18.58 & 0.08 & 18.53 & 0.04 & 18.55 & 0.07 & 18.57 & 0.05 & 18.55 & 0.03 & -5.89375 & -2.91056\\
36 & 18.48 & 0.10 & \ldots & \ldots & 18.48 & 0.10 & 18.47 & 0.10 & 18.48 & 0.06 & 14.73120 & -2.86056\\
37 & 18.59 & 0.10 & 18.51 & 0.10 & 18.53 & 0.10 & 18.54 & 0.10 & 18.54 & 0.05 & 7.10625 & -4.91889\\
38 & 18.64 & 0.10 & 18.60 & 0.15 & 18.65 & 0.10 & 18.59 & 0.10 & 18.62 & 0.05 & -12.76880 & -3.11889\\
39 & 18.61 & 0.10 & 18.52 & 0.06 & 18.63 & 0.06 & 18.54 & 0.10 & 18.58 & 0.04 & -5.64375 & -5.19389\\
\enddata

%% Include any \tablenotetext{key}{text}, \tablerefs{ref list},
%% or \tablecomments{text} between the \enddata and 
%% \end{deluxetable} commands

\tablenotetext{*}{Field used in the bar analysis.}
%% No \tablecomments indicated

%% No \tablerefs indicated

\caption{\small Distances to the 39 2MASS fields described in the text using the calibration of \autoref{sect:calib}. The distances are not corrected for reddening. Errors are determined from the smallest GLOESS smoothing window at which the TRGB peak remains at least twice the amplitude of any other response peak. In cases where there are multiple equal-amplitude peaks near the tip, half the spacing between the most separated peaks is taken to be the error. Differential positions are centered on RA: 05:23:34.5, Dec: -69:45:22, or, in degrees RA: 80.89375, Dec: -69.75611. }
\label{tab:2massdists}
\end{deluxetable*}

%1,  5,  6, 13, 14, 18, 19, 21, 23, 25, 27, 29, 30, 31, 32

%%% EXAMPLE DETECTIONS

%%% DISTANCE HISTOGRAMS

We note that the 2MASS TRGB detections, not the photometric zero-points, discussed in this section are independent of the detections that were used to establish the earlier provided calibration. Both use the slopes from Paper I, while the zero-point calibration uses the \citet{mac15} NISS bar observations, and this section use the 2MASS catalogs. Though expected, it is not a given that the two analyses agree to such a degree.

\section{Summary and Conclusions}
We have, together with a calibration of the slopes of the \(JHK\) TRGB from Paper I, presented a zero point calibration of the $JHK$ TRGB. Using this calibration we have also provided evidence for the back-to-front geometry of the LMC. These findings, in combination with a future independent calibration of the TRGB using \emph{Gaia}, indicate that the TRGB method at near-infrared wavelengths will provide a powerful tool for the measurement of extragalactic distances.

\section*{Acknowledgements} We thank the {\it Carnegie Institution for Science} and the {\it University of Chicago} for their continuing generous support of our long-term research into the expansion rate of the Universe. This research has made use of the NASA/IPAC Extragalactic Database (NED) and the {\it Infrared Science Archive} (IRSA) both of which are operated by the Jet Propulsion Laboratory, California Institute of Technology, under contract with the National Aeronautics and Space Administration. Direct support for this work was also provided by NASA through grant number HST-GO-13691.003-A from the Space Telescope Science Institute, which is operated by AURA, Inc., under NASA contract NAS 5-26555. Finally, we thank \citet{mac15} for making their near-infrared photometry freely available to the wider astronomical community. MGL was supported by a grant from the National Research Foundation (NRF) of Korea, funded by the Korean Government (MSIP) (NRF-2017R1A2B4004632).

\clearpage
\section*{Appendix} \label{sect:app}
For $J$ and $K$ magnitudes \cite{gor16} give:
\begin{align*}
M_J &= -5.67 -0.31~\mathrm{[Fe/H]} \\
M_K &= -6.98 -0.58~\mathrm{[Fe/H]}.
\end{align*}
\par\noindent
Combining/differencing these two equations leads to the
following metallicity-color relation
$$[Fe/H] = -4.85 + 3.70~(J-K)$$

Substituting $(J-K)$ for $[Fe/H]$ in their two preceding equations, and
re-centering them each at a fiducial color of $(J-K) = 1.00$ gives
\begin{align}
M_J &= -5.32 -1.15~[(J-K)-1.00]\\
M_K &= -6.32 -2.15~[(J-K)-1.00]
\end{align}
\par\noindent 
To make a homogeneous comparison with the presented calibration Equation 8 is entered into \autoref{tab:calcomp}.

% BIBLIOGRAPHY
\bibliographystyle{aasjournal}

%\bibliography{ms.bib}

\end{document}